\documentclass[prb, twocolumn, superscriptaddress, aps]{revtex4}
\usepackage{amsmath,amsfonts, amssymb, amsthm}
\usepackage{bm}
\usepackage{mathrsfs}
\usepackage{graphicx}
\usepackage{verbatim}
\usepackage{hyperref}
\usepackage{yfonts}
\usepackage{dsfont}
\usepackage{tikz}

\usetikzlibrary{external}
\newcommand{\di}{\mathrm{d}}

\newcommand{\ket}[1]{|#1\rangle}

\renewcommand{\ol}[1]{\overline{#1}}
\newcommand{\comments}[1]{}

\newcommand{\newsection}[1]{\section{#1}}
\newcommand{\mb}[1]{\mathbf{#1}}
\newcommand{\cohosub}[1]{\scalebox{0.72}{\textswab{#1}}}

\newcommand{\coho}[1]{\textswab{#1}}

\begin{document}

\title{Symmetry Fractionalization in Three-Dimensional $\mathbb{Z}_2$ Topological Order and Fermionic Symmetry-Protected Phases}
\author{Meng Cheng}
\affiliation{Station Q, Microsoft Research, Santa Barbara, California 93106-6105, USA}
\date{\today}

\begin{abstract}
	In two-dimensional topological phases, quasiparticle excitations can carry fractional symmetry quantum numbers. We generalize this notion of symmetry fractionalization to three-dimensional topological phases, in particular to loop excitations, and propose a partial classification for symmetry-enriched $\mathbb{Z}_2$ toric code phase. We apply the results to the classification of fermionic symmetry-protected topological phases in three dimensions.
\end{abstract}

\maketitle

Recent years have seen great advances in the classification and characterization of gapped quantum phases of matter.  At the most fundamental level, gapped quantum phases are characterized by topological properties of low-energy excitations, such as the fusion rules and braiding statistics of quasiparticles in two dimensions. In addition, if global symmetries are taken into account, a topological phase can split into several symmetry-enriched topological phases (SET)~\cite{Ying_2012, HungPRB2013, Lu_arxiv2013, BBCW, Yao_unpub}, distinguished by the interlay between symmetry and the topological degrees of freedom. In particular, it has been well-known that quasiparticle excitations can carry fractionalized quantum numbers of the global symmetry group (compared with quantum numbers carried by local excitations), and notable examples include Laughlin quasiholes in $\nu=\frac{1}{3}$ FQH state carrying $\frac{e}{3}$ electric charge, and spin-$1/2$ spinon excitations in a $\mathrm{SO}(3)$-symmetric $\mathbb{Z}_2$ spin liquid.
More recently, a rather complete understanding of global symmetry fractionalization in two-dimensional topological phases has been achieved~\cite{essin2013, BBCW, Tarantino_SET, TeoSET}, based on the notion of localized symmetry actions on quasiparticles. 
Intriguingly, it was found that certain symmetry fractionalization classes, albeit consistent with the underlying topological order, can not be realized in purely two-dimensional systems. Instead, they must exist on the surface of a three-dimensional symmetry-protected phase~\cite{senthil_3D, wang2013, Chen_arxiv2014, kapustin2014, ChoPRB2014} and in fact provide a non-perturbative characterization of the bulk SPT phase~\cite{senthil_3D, wang2014, Wang_PRB2014, metlitski2013, Chen_PRB2014, wang2013b, Fidkowski_PRX2013, WangPRB2014, Metlitski_arxiv2014}. 

In this work we will explore symmetry-enriched topological phases in 3D. Compared to 2D, the main difference is that 3D topological phases support extended loop-like excitations, the understanding of which remains rather limited. Furthermore, loops can also form nontrivial knots and links (or even more complicated branched structures), which further complicate the matter. We will present some partial results on classifying the symmetry transformations of both quasiparticles and loop excitations, which we generally refer to as symmetry fractionalization, lacking a better name.  
We focus on the simplest type of 3D topological phases, a $\mathbb{Z}_2$ toric code phase.
In a $\mathbb{Z}_2$ toric code there are two kinds of topologically nontrivial excitations: $\mathbb{Z}_2$ quasiparticle excitations $e$ and $\mathbb{Z}_2$ loop excitations $m$, which have mutual $\pi$ braiding phase. The $\mathbb{Z}_2$ charges can have bosonic or fermionic statistics, the latter of which is refered to as a fermionic toric code(FTC). The bosonic/fermionic $\mathbb{Z}_2$ toric code can be understood as a $\mathbb{Z}_2$ gauge theory coupled to bosonic/fermionic matter. 

Another motivation to study symmetry-enriched FTC is the intimate relation to 3D fermionic symmetry-protected(fSPT) phases. As previously shown in Ref. [\onlinecite{Cheng_fSPT}], gauging the fermion parity in a fSPT phase results in a symmetry-enriched $\mathbb{Z}_2$ topological order with fermionic charges, i.e. a $\mathbb{Z}_2$ FTC.  Although a fairly complete understanding of electronic topological insulators and superconductors (notice that electrons are Kramers doublets under the time-reversal symmetry) has been obtained~\cite{kitaev2009, schnyder2008, wang2014}, the seemingly simpler problem of classifying SPT phases with fermions transforming linearly has not been completely settled. Part of the reason is that in contrary to electronic topological insulators/superconductor, there are no nontrivial non-interacting 3D fSPT phases where fermions transform linearly under the symmetry. For example, if the symmetry is unitary and Abelian, for non-interacting systems one can always block diagonalize the single-particle Hamiltonian according to global symmetry quantum number, and the classification reduces to that of the class D within each block. It is known that class D has no nontrivial SPT phases, so we conclude that for non-interacting fermions there are no nontrivial SPTs with Abelian symmetry group. This argument obviously breaks down for interacting fermions, since the Hamiltonian can no longer be block diagonalized. A number of proposals for the classification of interacting fSPT phases have been put forward~\cite{GuWen, KapustinFSPT, Freed2014}, and the results are not completely consistent, in particular for $G=\mathbb{Z}_2^T$, namely time-reversal-invariant superconductors with $\mathcal{T}^2=1$. 

As an application of our results, we will interpret the Gu-Wen construction in 3D as symmetry fractionalization on flux loops of the $\mathbb{Z}_2$ fermion parity symmetry, and provide a physical argument that there are no nontrivial $\mathbb{Z}_2$ fSPT phases in 3D. We also propose a possible construction of a nontrivial fSPT phase with $\mathcal{T}^2=1$ symmetry.

The work is organized as follows. In Sec. \ref{sec:symfrac} we review the theory of symmetry fractionalization in two-dimensional topological phases, and derive the classification of symmetry fractionalization in 3D $\mathbb{Z}_2$ SET phases in two complimentary ways. We also present some examples of 3D $\mathbb{Z}_2$ SET phases obtained from partially gauging bosonic SPT phases. In Sec. \ref{sec:fspt} we apply our results to the classification of 3D fermionic SPT phases.

\newsection{Symmetry fractionalization in the $\mathbb{Z}_2$ topological phases}
\label{sec:symfrac}
First of all, we shall describe how symmetry can fractionalize in a 3D topological phase.  In particular, we will study how loop-like excitations transform under the symmetry, which is a new feature in 3D.

\subsection{Symmetry fractionalization in two-dimensional topological phases}
For pedagogical purpose, we first review symmetry fractionalization in two-dimensional SET phases, following the presentation in Ref. [\onlinecite{BBCW}].  Let $G$ be an on-site unitary symmetry group. The argument below applies to any SET phases where symmetries do not permute anyon types.
Let us prepare a state $\ket{\Psi_{\{a_j\}}}$ with $n$ well-separated quasiparticles, where the anyon type of the $j$th quasiparticle is $a_j$. (We use the convention where the ``vacuum'' topological charge is denoted $I$ and the ``topological charge conjugate'' of $a$ is denoted $\bar{a}$, which is the unique anyon type that can fuse with $a$ into vacuum.) We assume the overall topological charge is trivial $I$, so that the state can be created from the ground state by applying local operators.  We consider the global symmetry transformation $R_\mathbf{g}$ acting on $\ket{\Psi_{\{a_j\}}}$. Without loss of generality, we shall assume that the system transforms as a linear representation of $G$. Although anyons are non-local excitations, the local properties (i.e. local density matrix) of regions away from the positions of the anyons remain the same as those of the ground state.
Therefore the global symmetry transformation $R_\mathbf{g}$ should have the following decomposition (when the symmetries do not permute anyon types):
\begin{equation}
R_\mathbf{g}\ket{\Psi_{\{a_j\}}}=\prod_{j=1}^{n} U_\mathbf{g}^{(j)}\ket{\Psi_{\{a_j\}}}.
	\label{eq:sym_loc}
\end{equation}
Here, $U_\mathbf{g}^{(j)}$ is a local unitary operator whose nontrivial action is localized in the neighborhood of the $j$th quasiparticle.
 $U_\mathbf{g}^{(j)}$ only needs to projectively represent group multiplication:
\begin{equation}
U_\mathbf{g}^{(j)}U_\mathbf{h}^{(j)}\ket{\Psi_{\{a_j\}}}=\eta_{a_j}(\mathbf{g},\mathbf{h})U_\mathbf{gh}^{(j)}\ket{\Psi_{\{a_j\}}}
,
\label{eq:eta_def}
\end{equation}
where the projective phase $\eta_{a_j}(\mathbf{g},\mathbf{h}) \in \mathrm{U}(1)$ only depends on the topological properties localized in the neighborhood of the $j$th quasiparticle, which is just the anyon type $a_j$. Since $R_{\bf g}$ is a linear representation, the projective phases must satisfy the condition $\prod_j \eta_{a_j}(\mathbf{g}, \mathbf{h})=1$.
 This condition is equivalent to the condition that $\eta_a(\mathbf{g},\mathbf{h})\eta_b(\mathbf{g},\mathbf{h})=\eta_c(\mathbf{g},\mathbf{h})$ whenever the topological charge $c$ is a permissible fusion channel of the topological charges $a$ and $b$. It follows that $\eta_a(\mathbf{g},\mathbf{h})$ must take the form~\cite{BBCW}
\begin{equation}
\eta_a(\mathbf{g},\mathbf{h})=M_{a, \cohosub{w}({\bf g},{\bf h})},
\label{eqn:etaasbraiding}
\end{equation}
where $\coho{w}(\mathbf{g},\mathbf{h})$ is an Abelian anyon, and $M_{a, \cohosub{w}({\bf g},{\bf h})}$ is the mutual braiding statistics between anyons $a$ and $\coho{w}({\bf g},{\bf h})$.

Associativity of the localized operators gives the condition
\begin{equation}
\eta_a (\mb{g,h}) \eta_a (\mb{gh,k})= \eta_a (\mb{g,hk}) \eta_a (\mb{h,k})
,
\end{equation}
for all $a$, which translates into
\begin{equation}
\coho{w} (\mb{g,h}) \times \coho{w} (\mb{gh,k}) = \coho{w} (\mb{g,hk}) \times \coho{w} (\mb{h,k})
.
\end{equation}
This is precisely the $2$-cocycle condition. Therefore $\coho{w}(\mb{g,h})$ are $2$-cocycles valued in the group of Abelian anyons $\mathcal{A}$ with group multiplication given by the fusion rules. One can further show that there are ambiguities of $\coho{w}(\mb{g,h})$ from redefinitions of the localized symmetry transformations, exactly translated into $2$-coboundaries valued in $\mathcal{A}$~\cite{BBCW}. Therefore, symmetry fractionalization is classified by the cohomology group $\mathcal{H}^2[G, \mathcal{A}]$. Given a $2$-cocycle $\coho{w}$, the local projective phases $\eta_a (\mb{g,h})$ are completely determined through Eq. \eqref{eqn:etaasbraiding} for all anyon types.

\subsection{Symmetry fractionalization of $\mathbb{Z}_2$ charges}
We now turn to the study of symmetry fractionalization in 3D $\mathbb{Z}_2$ SET phases. For $\mathbb{Z}_2$ charge excitations, the previous argument in 2D can be generalized straightforwardly. The global symmetry transformation can be decomposed to local unitaries on the $\mathbb{Z}_2$ charges, which satisfy the group multiplication law projectively. Due to the $\mathbb{Z}_2$ fusion rules, the projective phases $\eta(\mb{g,h})$ only take values $\pm 1$, which should now be associated with the braiding of the charge with  $\mathbb{Z}_2$ flux loops. The classification is given by $\mathcal{H}^2[G, \mathbb{Z}_2]$. We will denote the fractionalization class by $[\coho{w}_e]\in \mathcal{H}^2[G, \mathbb{Z}_2]$.

\subsection{Symmetry action on $\mathbb{Z}_2$ flux loops}
Next we consider the symmetry action on loop excitations.  Now the situation becomes much more complicated, since loops can link with each other, and it seems impossible to exhaust all possible links.  For simplicity, we assume that the fractionalization class $[\coho{w}_e]$ of the $\mathbb{Z}_2$ charge is trivial. 

\subsubsection{Dimensional reduction}
It is useful to visualize a flux loop excitation to actually bound an invisible membrane. In fact, one way to understand the ground state wavefunction of a $\mathbb{Z}_2$ toric code in 3D is to view it as a superposition of fluctuating membrane configurations, as a generalization of the ``string-net'' condensation picture in 2D~\cite{LevinWen}. Alternatively, one can create a flux loop excitation by applying a membrane operator to the ground state.

\begin{figure}
	\includegraphics[width=\columnwidth]{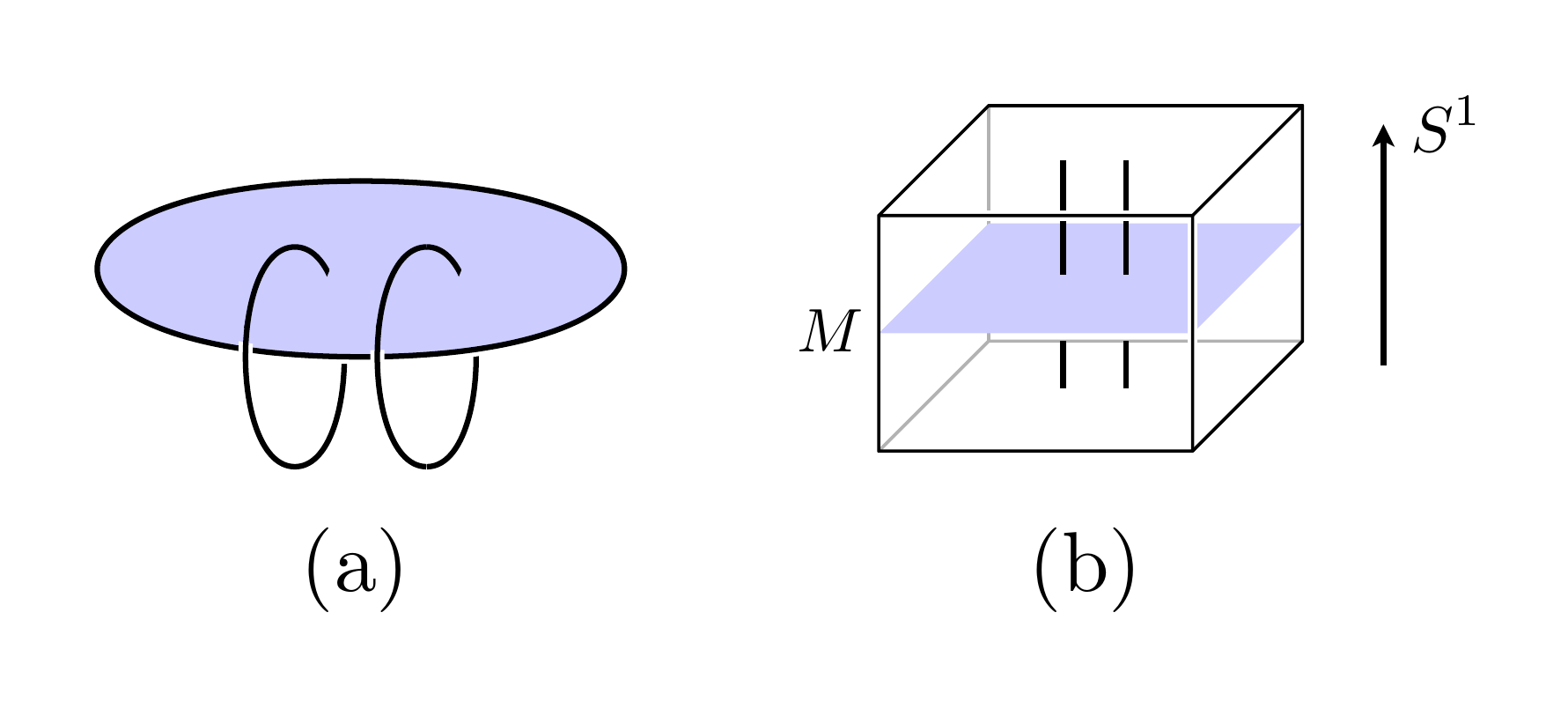}
\caption{ (a) Linked flux loops. (b) Dimensional reduction of the system on $M\times S^1$. The plane represents the flux membrane along a 2D surface $M$, and flux loops along $S^1$ intersect the membrane, which can be thought as $m$ anyons in a 2D toric code on $M$.}
\label{dimreduction}
\end{figure}

Heuristically, one may imagine that the membrane can be thought as a 2D topological phase with interesting symmetry transformations. To make it precise, we perform a dimensional reduction in the following way: consider the $\mathbb{Z}_2$ topological order on a spatial manifold $M\times S^1$, where $M$ is a closed surface. We choose the ground state to have a $\mathbb{Z}_2$ flux threading the ``hole'' of $S^1$, i.e. when a $\mathbb{Z}_2$ charge goes around $S^1$ it sees a $-1$ Berry phase. Equivalently, one can also imagine that a $\mathbb{Z}_2$ flux membrane (created by applying the membrane operator along $M$, for example) intersects $S^1$. For an illustration we refer to Fig. \ref{dimreduction}(b).  If the radius of $S^1$ is much shorter than the size of $M$, we can view the whole system as a 2D $\mathbb{Z}_2$ toric code: namely, flux loops wrapping around $S^1$ become the $\tilde{m}$ anyons (we use $\tilde{}$ to distinguish the labels in the 2D SET obtained from dimensional reduction from their 3D counterparts), and the $\mathbb{Z}_2$ charges in the bulk naturally descend to the $\tilde{e}$ anyons. Essentially,  using this dimensional reduction procedure we can study, in a very precise sense, the 2D topological phase ``attached'' to a $\mathbb{Z}_2$ flux membrane, and the corresponding flux loop become the edge of the 2D topological phase (See Fig. \ref{dimreduction}(a) ). This observation was used in Ref. [\onlinecite{Levin_PRL2014}] to compute loop braiding statistics (see also Ref. [\onlinecite{chen_3d}]).

Symmetry fractionalization in the dimensionally reduced 2D toric code is classified by $\mathcal{H}^2[G, \mathbb{Z}_2\times\mathbb{Z}_2]=\mathcal{H}^2[G, \mathbb{Z}_2]\times \mathcal{H}^2[G, \mathbb{Z}_2]$. Translating back to 3D, we see that the fractionalization of $\tilde{e}$ is obviously the same as $[\coho{w}_e]$. The fractionalization of $\tilde{m}$, denoted by $[\coho{w}_m]\in \mathcal{H}^2[G, \mathbb{Z}_2]$, has a more nontrivial 3D interpretation. Because the $\tilde{m}$ anyon in this dimensionally reduced setting is in fact the intersection of a flux loop with the membrane bounded by the other flux loop, we will refer to $[\coho{w}_m]$ as the symmetry fractionalization of loop-membrane intersections for linked loops. Because of our simplifying assumption that $[\coho{w}_e]=[1]$, we have $\coho{w}_m(\mb{g,h})\in \{I, e\}$.

With this point of view, it is quite natural that the flux loops themselves can carry edge states of a 2D SPT phase, classified by $\mathcal{H}^3[G, \mathrm{U}(1)]$~\cite{ChenPRB2013, Chen_science}. To be compatible with the $\mathbb{Z}_2$ nature of the flux loop, only the order-$2$ elements of the cohomology group $\mathcal{H}^3[G, \mathrm{U}(1)]$ are allowed. Intuitively, one can imagine the 2D SPT phase is attached to the membrane bounded by the flux loop~\cite{Chen_NC2014}. However, we will see below that this is not the complete story.

In summary, the dimensional reduction procedure allows one to classify the symmetry fractionalization on flux loops via the theory of 2D symmetry-enriched topological phases.

\subsubsection{Fractionalization on  flux loops}
Let us now examine directly the symmetry action on a flux loop, without appealing to the dimensional reduction argument. To completely define a single loop excitation, we also need to specify whether there are any $\mathbb{Z}_2$ charges attached to the loop. We will assume that the loop is ``neutral'', in the sense that overall there are no $\mathbb{Z}_2$ charges attached. This can be operationally defined as that the loop can be continuously shrinked to a local point-like excitation.

Following again the idea of symmetry localization, one can decompose the symmetry transformation on a global state with several flux loops present, to unitaries supported on the neighborhood of the flux loops. We shall denote these loop-like operators by $U_\mb{g}$ (labels for different flux loops are kept implicit for now). To characterize $U_\mb{g}$, one may first want to check whether they form  linear representations or not when acting on the state. For  a neutral loop that can be continuously deformed to a local excitation, one may naively assume that corresponding $U_\mb{g}$ must form a linear representation of $G$, i.e. $U_\mb{g}U_\mb{h}=U_\mb{gh}$.  When interpreted at the level of action on the state, the assumption is actually correct if loops do not link. However, as we will argue below, to account for more general situations, one should actually write 
\begin{equation}
	U_\mb{g}U_\mb{h}=\mathcal{W}(\mb{g,h})U_\mb{gh}.
	\label{}
\end{equation}
Here $\mathcal{W}$ is a closed ``string'' operator that acts trivially on the ground state. In this case, $\mathcal{W}$ is an $e$ charge transported along a closed path, or equivalently creates a pair of $e$ charges and separates them along the given path. In fact, it is not hard to see that the form of $\mathcal{W}$ is completely fixed by the $[\coho{w}_m]$ class: $\mathcal{W}(\mb{g,h})=\mathcal{W}_{\cohosub{w}_m(\mb{g,h})}$. To illustrate, we present a concrete example of a similar result for the edge of 2D SET phase in Appendix \ref{sec:2dsetedge}.

To further diagnose the symmetry action, we follow an elegant reduction procedure introduced in Ref. [\onlinecite{ElseSPT}],  which classifies SPT phases according to the anomalous symmetry actions on the boundary. First we review the argument in the context of 2D SPT phases. We will use the same notation $U_\mb{g}$ to denote the symmetry transformation on the edge of a 2D SPT. In this case, $U$ should form a linear representation of $G$, i.e. $U_\mb{g}U_\mb{h}=U_\mb{gh}$. We now restrict $U_\mb{g}$ to an open interval ${I}$ with end points $a$ and $b$. The restriction is denoted by $U_\mb{g}^I$. $U_\mb{g}^I$ does not have to form a linear representation, due to the ambiguities at the end points. In other words, they can be ``projective":
\begin{equation}
	U_\mb{g}^IU_\mb{h}^I=\Omega(\mb{g,h})U_\mb{gh}^I.
	\label{}
\end{equation}
But $\Omega(\mb{g,h})$ is generally not  merely a  phase, instead a unitary operator localized on the two end points. The associativity of $U_\mb{g}^I$ implies
\begin{equation}
	\Omega(\mb{g,h})\Omega(\mb{gh,k})=^{U^I_\mb{g}}\!\Omega(\mb{h,k})\Omega(\mb{g,hk}),
	\label{eqn:omega_asso}
\end{equation}
where $^U\Omega$ is a shorthand notation for $U\Omega U^{-1}$.

To complete the reduction, we decompose $\Omega$ into local unitaries at the end points, which can always be done on the edge of a 2D SPT. We consider the local unitary at the end point $a$ denoted by $\Omega_a$. It now satisfies the associativity relation only up to a phase:
\begin{equation}
	\Omega_a(\mb{g,h})\Omega_a(\mb{gh,k})=\nu(\mb{g,h,k})^{U^I_\mb{g}}\Omega_a(\mb{h,k})\Omega_a(\mb{g,hk}).
	\label{eqn:omega} 
\end{equation}
It is straightforward to show that $\nu(\mb{g,h,k})$ is a $3$-cocycle (for a proof, see Ref. [\onlinecite{ElseSPT}] and Appendix \ref{sec:3cocycle}).

Therefore we have defined a procedure to extract a $3$-cocycle, $\nu(\mb{g,h,k})$ from the restriction of symmetry actions on the edges. There are two places where ambiguities may arise: a)
the choice of the restriction $U\rightarrow U^I$ (i.e. modifying the restriction by a local unitary near the two end points). One can easily show that this choice does not change $\nu$ at all. b) When we restrict $\Omega$ to $\Omega_a$ there is a phase ambiguity $\Omega_a(\mb{g,h})\rightarrow \beta(\mb{g,h})\Omega_a(\mb{g,h})$, which precisely translates into a $3$-coboundary for $\nu(\mb{g,h,k})$. Therefore, the symmetry action on the edge is classified by $\mathcal{H}^3[G, \mathrm{U}(1)]$.

We now generalize this argument to flux loops in 3D. The main obstacle in applying this reduction procedure is that the operator $\Omega$ is \emph{not} localized near the end points, due to the presence of $\mathcal{W}$.  We can still restrict $U_\mb{g}$ to an interval and define $\Omega$, however at best we can hope to represent $\Omega(\mb{g,h})$ as
\begin{equation}
	\Omega(\mb{g,h})\sim\Omega_a(\mb{g,h})\Omega_b(\mb{g,h})\mathcal{W}^I_{\cohosub{w}(\mb{g,h})},
\end{equation}
where $\Omega_{a,b}$ are local unitaries.  It is not clear how to do this decomposition canonically. Formally, one can attempt to ``localize'' both sides of Eq. \eqref{eqn:omega_asso} to the end points and then calculate $\nu$, however this is still problematic. To illustrate, we 
consider redefining $U_\mb{g}=\tilde{U}_\mb{g}\mathcal{W}_{\cohosub{z}(\mb{g})}$, and
\begin{equation}
	\tilde{\Omega}(\mb{g,h})=\mathcal{W}_{\cohosub{z}(\mb{gh})}^{-1}\mathcal{W}_{\cohosub{z}(\mb{g})}\, ^{U_\mb{g}^I}\mathcal{W}_{\cohosub{z}(\mb{h})}\Omega(\mb{g,h}).
	\label{eqn:Omega1}
\end{equation}
This redefinition corresponds to modifying $\coho{w}$ by a coboundary $\coho{z}$.
We will calculate how the redefinition in Eq. \eqref{eqn:Omega1} affects the $3$-cocycle $\nu$. We can naively plug Eq. \eqref{eqn:Omega1} into the associativity relation, and find $^{U_\mb{gh}^I}\mathcal{W}_{\cohosub{z}(\mb{k})}$ on the left-hand side, and $^{ U_\mb{g}^IU_\mb{h}^I}\mathcal{W}_{\coho{z}(\mb{k})}$ on the right-hand side. Recall that $\mathcal{W}_{\cohosub{z}(\mb{k})}$ is the open string operator that create a $\coho{z}(\mb{k}), \ol{\coho{z}(\mb{k})}$ pair, so one can formally decompose $^{U_\mb{gh}^I}\mathcal{W}_{\cohosub{z}(\mb{k})}$ into local unitaries at the end points $a$ and $b$ where $\coho{z}(\mb{k})$ and $\ol{\coho{z}(\mb{k})}$ sit. Comparing the two sides, we see that the result is exactly a phase $\eta_{\cohosub{z}(\mb{k})}(\mb{g,h})$. Thus $\nu(\mb{g,h,k})$ is modified:
\begin{equation}
	\nu(\mb{g,h,k})\rightarrow \nu(\mb{g,h,k}) \eta_{\cohosub{z}(\mb{k})}(\mb{g,h}),
	\label{eqn:omega_ambiguity}
\end{equation}
under the redefinition in Eq. \eqref{eqn:Omega1}~\footnote{The same argument applies to the edge of a 2D SET phase. We should however notice that for general anyon models, there may be additional contributions from braiding.}. Therefore, whenever $[\coho{w}_m]$ is nontrivial, it is possible that one can choose $\coho{z}(\mb{k})$ to cancel $\nu(\mb{g,h,k})$. Suggested by this heuristic argument, in order to eliminate this ambiguity, we only apply the reduction procedure when the $[\coho{w}_m]$ class is trivial.


From now on we focus on the $[\coho{w}]\equiv [1]$ case, 
 and work in a canonical gauge where $\coho{w}\equiv I$ to get rid of the string operators in $\Omega_\mb{g}$ all together.
The operator $\tilde{\Omega}(\mb{g,h})$ now can be restricted to the end points to give a local unitary, which then satisfies associativity relation up to a $3$-cocycle, as done in Eq. \eqref{eqn:omega}.

So far the argument is parallel to the classification of anomalous symmetry actions on the edge of a 2D SPT phase. Similar to what happens for the charges, we now take into account the fusion rules of flux loops. Namely, two flux loops should fuse to the vacuum. If we have two flux loops sitting on top of each other, by applying local unitaries one can completely annihilate these two flux loops, and bring the system back to the ground state. This fact can be used to show that $\nu(\mb{g,h,k})=\pm 1$:
On one hand, treating the two flux loops, labeled $1$ and $2$, as one loop excitation, we have $U_\mb{g}={U_\mb{g}^{(1)}}{U_\mb{g}^{(2)}}$, and similarly for $U_\mb{g}^I$. Therefore the $3$-cocycle one would extract for the combined loop excitation must be $[\nu(\mb{g,h,k})]^2$. On the other hand, if we fuse the two flux loops (by applying local unitaries), the system returns to the ground state, and thus $U_\mb{g}$ is trivially the restriction of the global symmetry operator, which is just a product of local on-site symmetry transformations. This implies that $\Omega(\mb{g,h})\equiv \mathds{1}$. The local unitaries applied to fuse the two loops can at most change $U_\mb{g}^I$ by some local unitaries, and it has been shown that this does not affect the $3$-cocycle~\cite{ElseSPT}.
It is then immediately follows that the topological classification of $U_\mb{g}$ is given by $\mathcal{H}^3[G, \mathbb{Z}_2]$. We will denote the cohomology class by $[\nu]$, and will generally refer to $[\nu]$ as the fractionalization class of flux loops.

We have so far assumed $G$ is a unitary symmetry group. The argument can also be generalized to anti-unitary symmetries, and the details can be found in Appendix \ref{sec:3cocycle}.

Notice that with the assumption of trivial $\coho{w}$, we have restricted ourselves to the phases where the gauge charges transform linearly under the symmetry. We believe all such phases can be obtained by gauging the $\mathbb{Z}_2$ global symmetry in a $\mathbb{Z}_2\times G$ SPT phase.

\subsection{Examples}
We now study several examples of 3D symmetry-enriched $\mathbb{Z}_2$ topological phase, from partially gauging the $\mathbb{Z}_2$ symmetry in a $\mathbb{Z}_2\times G$ SPT phase. A similar construction of SET phases in two dimensions was considered in Ref. [\onlinecite{Ying_2012}]. When the $\mathbb{Z}_2$ subgroup is gauged, the result is a $G$-enriched $\mathbb{Z}_2$ toric code. It is easy to see that the $\mathbb{Z}_2$ charges transform linearly under $G$, since the symmetry group is a direct product of $\mathbb{Z}_2$ and $G$. We are particularly interested in the SPT phases that are protected only when both $\mathbb{Z}_2$ and $G$ are present.  For those SPT phases, we can argue that the $\mathbb{Z}_2$ flux loops must transform nontrivially under $G$, i.e. it is impossible to condense the flux loops without breaking $G$. Otherwise, if we condense the flux loops while preserving the $G$ symmetry, all excitations that are charged under $\mathbb{Z}_2$ are confined. What then remains is essentially just a SPT phase with $G$ symmetry, contradicting with the assumption that both $\mathbb{Z}_2$ and $G$ are needed for the phase to be nontrivial.

Before we go to the examples, it is instructive to first calculate $\mathcal{H}^4[\mathbb{Z}_2\times G, \mathrm{U}(1)]$ explicitly using K\"unneth formula. For a unitary $G$ we obtain 
\begin{equation}
	\mathcal{H}^4[\mathbb{Z}_2\times G, \mathrm{U}(1)]=\mathcal{H}^3[G, \mathbb{Z}_2]\times \mathcal{H}^1[G, \mathbb{Z}_2].
	\label{}
\end{equation}

\subsubsection{$G=\mathbb{Z}_N$}
First we consider $G=\mathbb{Z}_N$, and we will denote the generator of $\mathbb{Z}_N$ by $\mb{g}$.  We have $\mathcal{H}^3[G, \mathrm{U}(1)]=\mathbb{Z}_N$ while $\mathcal{H}^3[G, \mathbb{Z}_2]=\mathbb{Z}_{(2,N)}$. Focusing on even $N$, we can easily see that there is an ``order two'' 2D SPT phase protected by $\mathbb{Z}_N$.  Physically, the nontrivial $\mathcal{H}^3[G, \mathbb{Z}_2]$ fractionalization class means that there are gapless modes on flux loops which have the same symmetry transformation as that of the edge of a 2D $\mathbb{Z}_N$ SPT phase. One can construct a wavefunction in the following way: first recall that the ground state wavefunction of a 3D $\mathbb{Z}_2$ order can be intuitively understood as a superposition of fluctuating $\mathbb{Z}_2$ membranes. Now we decorate the membranes with a 2D $\mathbb{Z}_N$ SPT state. In this example, the symmetry fractionalization on flux loops can be detected by introducing $\mathbb{Z}_N$ fluxes as extrinsic defects, and examine the three-loop braiding statistics $\Theta_{\mb{gg},m}$~\cite{Levin_PRL2014, WangLevinPRB, Ran_PRX2014} (for the definition of $\Theta$ we refer to Ref. [\onlinecite{WangLevinPRB}]). The nontrivial class corresponds to $\Theta_{\mb{gg},m}=\pi$.

A different SET phase is characterized by a fractional $\mathbb{Z}_N$ charge for the loop-membrane intersection. The corresponding second cohomology is $\mathcal{H}^2[\mathbb{Z}_N, \mathbb{Z}_2]=\mathbb{Z}_{(2,N)}$. Again one can detect this fractionalization class the three-loop braiding phase $\Theta_{\mb{g}m,m}$.

Naively one may posit the existence of a third nontrivial SET where both characteristics are present, by gauging a $\mathbb{Z}_2\times\mathbb{Z}_N$ SPT phase that is the ``sum'' of the previous two.  Quite counter-intuitively this is not the case, for the following reason: as we have emphasized, when a nontrivial fractionalization class for loop-membrane intersection $[\coho{w}_m]$ is present, the argument for the $\mathcal{H}^3[G, \mathbb{Z}_2]$ classification does not apply anymore. In this case with a unitary Abelian $G$, one may still attempt to define the symmetry fractionalization for a single flux loop by the corresponding three-loop braiding phase $\Theta_{\mb{gg},m}$.  However, this proposal still fails, since $\mb{g}$ is an extrinsic symmetry defect loop and one is allowed to attach any (intrinsic) excitation to $\mb{g}$~\footnote{We thank Xie Chen for emphasizing the importance of charge and flux loop attachment on defect loops.}. We then define $\mb{g}'=\mb{g}\times m$, and it is straightforward to see that $\Theta_{\mb{g}'\mb{g'},m}=\pi, \Theta_{\mb{g}'m,m}=\pi$. Therefore we conclude that two $\mathbb{Z}_2$ SET phases with (1) $\Theta_{\mb{gg},m}=0, \Theta_{\mb{g}m,m}=\pi$ and (2) $\Theta_{\mb{gg},m}=\pi, \Theta_{\mb{g}m,m}=\pi$ are topologically equivalent. This ``collapse'' is also consistent with Eq. \eqref{eqn:omega_ambiguity}, since for $G=\mathbb{Z}_2$ the nontrivial $[\nu]$ class is fully determined by $\nu(\mb{g,g,g})=-1$, which can be cancelled if we choose $\coho{z}(\mb{g})=m$ in Eq. \eqref{eqn:omega_ambiguity}. We notice that a completely parallel result has been known for $\mathbb{Z}_N$ symmetry-enriched $\mathbb{Z}_2$ toric code phase in two dimensions~\cite{Metlitski_unpub, BBCW, Burnell_unpub}.

In summary, we have found that when $[\coho{w}_e]=[1]$, there are three $\mathbb{Z}_2$ SET phases characterized by fractionalization on flux loops.

\subsubsection{$G=\mathbb{Z}_2\times\mathbb{Z}_2$}
Now we move on to a slightly more complicated case of $G=\mathbb{Z}_2^X\times\mathbb{Z}_2^Y$. The two types of symmetry fractionalization are classified by $\mathcal{H}^3[G, \mathbb{Z}_2]=\mathbb{Z}_2^4$ and $\mathcal{H}^2[G, \mathbb{Z}_2]=\mathbb{Z}_{2}^3$, respectively. Notice that $\mathcal{H}^3[G, \mathrm{U}(1)]=\mathbb{Z}_2^3$.

We first consider the $[\coho{w}_m]$ fractionalization class associated with loop-membrane intersections. Physically we can characterize them by the three-loop braiding processes $\Theta_{\mb{g}m,m}$ and $\Theta_{m\mb{gh},m}$ where $\mb{g,h}\in G$. The former detect fractional charges on the intersection, while the latter detect projective representations of $G$. Since we only consider those $\mathbb{Z}_2$ SETs that can be obtained from gauging $\mathbb{Z}_2\times G$ SPT phases~\cite{WangLevinPRB}, the $\Theta_{m\mb{gh},m}$ must vanish. In other words, the loop-membrane intersection can not carry a projective representation of $G=\mathbb{Z}_2\times \mathbb{Z}_2$. This is likely to hold more generally~\cite{chen_private}. 

Let us turn to the $\mathcal{H}^3$ fractionalization. To characterize those fractionalization classes we consider the three-loop braiding statistics of $G$ defect loops. In particular, we need to look at braiding of $G$ symmetry defects linked to a base $m$ loop, and there are three invariants: $\Theta_{XX,m}, \Theta_{YY, m}, \Theta_{XY,m}$ corresponding exactly to the three generators of $\mathcal{H}^3[G, \mathrm{U}(1)]=\mathbb{Z}_2^3$. We see that interestingly we miss one extra generator in $\mathcal{H}^3[G, \mathbb{Z}_2]=\mathbb{Z}_2^4$.

So far we have been able to match the symmetry fractionalization on flux loops with the following three-loop braiding phases: $\Theta_{Xm,m}, \Theta_{Ym,m}, \Theta_{XX,m}, \Theta_{YY, m}, \Theta_{XY,m}$. Compared to the complete list of three-loop braiding invariants involving $m$, the missing one is $\Theta_{mX,Y}$ ($\Theta_{mY,X}$ is related to $\Theta_{mX,Y}$ and $\Theta_{XY, m}$ by $\Theta_{Xm,Y}+\Theta_{Ym,X}+\Theta_{XY,m}=0$), which does not have a ``2D'' interpretation from dimensional reduction to the $m$ membrane. This strongly suggests that the generator of $\mathcal{H}^3[G, \mathbb{Z}_2]$ which is a trivial $\mathrm{U}(1)$ cocycle should be matched with $\Theta_{Xm,Y}=\Theta_{Ym,X}=\pi$ (while all other three-loop braiding phases are zero).



\subsubsection{$G=\mathbb{Z}_2^T$}
\label{sec:z2T}
For $G=\mathbb{Z}_2^T$, the classification becomes $ \mathcal{H}^3[G, \mathbb{Z}_2]=\mathbb{Z}_2$, but notice $\mathcal{H}^3[G, \mathrm{U}(1)]=\mathbb{Z}_1$, namely there are no 2D $\mathbb{Z}_2^T$ SPT phases. Therefore this is a case that is intrinsically ``3D''. We now argue that both two generators can be realized by partially gauging a $\mathbb{Z}_2\times\mathbb{Z}_2^T$ SPT. For later convenience, we denote $\mathbb{Z}_2\times\mathbb{Z}_2^T=\{1, \mb{g}, \mathcal{T}, \mb{g}\mathcal{T}\}$. Applying the K\"unneth formula~\cite{ChenPRB2013}:
\begin{equation}
	\mathcal{H}^4[\mathbb{Z}_2\times\mathbb{Z}_2^T, \mathrm{U}_\mathcal{T}(1)]=\mathbb{Z}_2^3.
	\label{}
\end{equation}
Among the three generators, one of them is obviously the 3D bosonic SPT protected just by $\mathbb{Z}_2^T$, which is not interesting for our purpose. The other two generators need both $\mathbb{Z}_2$ and $\mathbb{Z}_2^T$ symmetries to be nontrivial.  One of $\mathbb{Z}_2$ factors corresponds to nontrivial $[\coho{w}_m]$ fractionalization class in $\mathcal{H}^2[\mathbb{Z}_2^T, \mathbb{Z}_2]=\mathbb{Z}_2$ after gauging the $\mathbb{Z}_2$ symmetry. Physically, the nontrivial $[\coho{w}_m]$ means the loop-membrane intersection carries a Kramers doublet. We notice that $[\coho{w}_m]$ can be extracted from applying slant products with respect to $\mathbb{Z}_2$ twice to the $4$-cocycle.

The other $\mathbb{Z}_2$ factor (with $[\coho{w}_m]=[1]$) in the classification is more subtle, since the nontrivial action can not be detected via dimensional reduction. 
The wavefunction of such a state can be constructed in a similar way as the $G=\mathbb{Z}_2$ case: we introduce to the system two sets of Ising spins, say $A$ and $B$, and time-reversal symmetry flips $B$ spins (in addition to the global complex conjugation). Thus for each given configuration, $B$ spins form time-reversal domain walls. Let $N(\sigma_A, \sigma_B)$ denotes the number of intersections between the membranes of $A$ spins and the time-reversal domain walls in $\sigma_B$. The wavefunction is then
\begin{equation}
	\ket{\psi}=\sum_{\{\sigma_A,\sigma_B\}}(-1)^{N(\sigma_A,\sigma_B)}\ket{\sigma_A,\sigma_B}.
	\label{}
\end{equation}
We conjecture that \emph{this state exhibits symmetry fractionalization on flux loops, given by the nontrivial cohomology class in $\mathcal{H}^3[\mathbb{Z}_2^T, \mathbb{Z}_2]$}. We notice that the corresponding SPT phase before gauging can also be understood as decorating 2D $\mathbb{Z}_2$ SPT state to the $\mathbb{Z}_2^T$ domain walls.

\section{Fermion SPT Phases and $\mathbb{Z}_2$ FTC}
\label{sec:fspt}
We now turn to the other class of $\mathbb{Z}_2$ topological order, where the $\mathbb{Z}_2$ charge is fermionic. The analysis of the symmetry action on $\mathbb{Z}_2$ flux loops still applies. We will focus on the cases where the fermionic $\mathbb{Z}_2$ charges transform linearly under the symmetry group $G$. They can be obtained from gauging the fermion parity symmetry in fSPT phases, and hence classifying the SET orders provides a (partial) classification of fSPT phases.

For interacting fermionic SPT phases, Gu and Wen developed a group super-cohomology construction~\cite{GuWen}. 
It is instructive to review briefly the Gu-Wen classification. Gu and Wen in a true \emph{tour de force}, constructed a topologically-invariant partition function for fermions in any $d$ spatial dimensions. The data that enters the construction is a $d$-cocycle from $\mathcal{H}^d[G, \mathbb{Z}_2]$. Crucially, in order to define the partition function, one needs to check that certain obstruction class, living in $\mathcal{H}^{d+2}[G, \mathrm{U}(1)]$ vanishes. The obstruction class can be computed from the $d$-cocycle using a cohomology operation known as the Steenrod square. In $d=2$, the obstruction has been re-derived from the classification of SET phases obtained from gauging the fermion parity~\cite{Cheng_fSPT}, and the cohomology class from $\mathcal{H}^2[G, \mathbb{Z}_2]$ is interpreted as the symmetry fractionalization class of the $\mathbb{Z}_2$ fermion parity flux. 

Besides the Gu-Wen classification, another general classification scheme has been proposed in Ref. [\onlinecite{KapustinFSPT}] based on bordism theory.  The results of the two classifications agree for spatial dimensions $<3$. In three dimensions, both classification schemes obtain no nontrivial fSPT phases for $G=\mathbb{Z}_2$ (and easily generalized to $G=\mathbb{Z}_N$). However, for $G=\mathbb{Z}_2^T$, Gu and Wen obtained a $\mathbb{Z}_2$ classification while the bordism approach still gives no nontrivial phases.

Based on our results, we conjecture that in three dimensions, the cohomology class from $\mathcal{H}^3[G, \mathbb{Z}_2]$ in the Gu-Wen construction can be understood as the $\mathcal{H}^3$ fractionalization classes of the $\mathbb{Z}_2$ fermion parity flux loops. In addition, there are also $\mathcal{H}^2$ fractionalization classes for loop-membrane intersections, which are not considered in the Gu-Wen construction.

\section{Triviality of $\mathbb{Z}_2$ Fermion SPT}
In this section we give a physical argument that nontrivial 3D fSPT phases do not exist for $G=\mathbb{Z}_2$. If such a fSPT exist, by gauging the fermion parity one can obtain $\mathbb{Z}_2$ FTC. The nontrivialness of the fSPT has to be reflected in the symmetry action on the flux loops, otherwise if the flux loops are completely trivial we can then proliferate the flux loops and confine all the fermionic $\mathbb{Z}_2$ charges without breaking the symmetry. The result would be a purely bosonic SPT phase.

Based on our conjecture, if such nontrivial $\mathbb{Z}_2$ fSPT phases exist, they must be characterized by $[\nu]\in\mathcal{H}^3[\mathbb{Z}_2, \mathbb{Z}_2]$ and $[\coho{w}_m]\in \mathcal{H}^2[\mathbb{Z}_2, \mathbb{Z}_2]$ in the gauged $\mathbb{Z}_2$ FTC. Notice that for $G=\mathbb{Z}_2$, both $[\nu]$ and $[\coho{w}_m]$ can be understood from the dimensional reduction procedure.

First we rule out the possibility of a nontrivial $[\nu]$ class. We will show that one can construct such a $\mathbb{Z}_2$ FTC at the boundary of a 4D $\mathbb{Z}_2$ SPT phase, and thus this SET is anomalous. We follow the field-theoretical construction in Ref. [\onlinecite{Bi_unpub}], and describe the bulk 4D $\mathbb{Z}_2$ SPT phase by a $\mathrm{O}(6)$ nonlinear $\sigma$ model(NLSM) with a topological $\Theta$-term~\cite{senthil_3D, XuNLSM}:
\begin{multline}
	S=\int\di^4x\di\tau\,\frac{1}{g}(\partial_\mu\mathbf{n})^2\\
	+\frac{i\Theta}{\Omega_5}\varepsilon_{abcdef}n^a\partial_x n^b \partial_y n^c \partial_z n^d\partial_w n^e \partial_\tau n^f.
	\label{}
\end{multline}
Here $|\mathbf{n}|=1$ is a six-component unit vector and $\Omega_5$ is the volume of the five dimensional unit sphere.

It is well-known that the topological term is quantized to $\Theta\mathbb{Z}$ on a closed manifold, and therefore does not contribute to the partition function for $\Theta=2\pi$. The bulk spectrum is thus completely determined by the gradient term. When $g$ is sufficiently large, the system has a gapped and non-degenerate bulk. In order to describe a nontrivial SPT phase, we assign the following $\mathbb{Z}_2$ symmetry transformation to $\mathbf{n}$:
\begin{equation}
	\mathbb{Z}_2: \mathbf{n}\rightarrow -\mathbf{n}.
	\label{}
\end{equation}
In fact, it is useful to first enhance the symmetry to $\mathrm{SO}(3)_1\times \mathrm{SO}(3)_2$, which rotate $\mathbf{N}=(n_1,n_2,n_3)$ and $\mathbf{M}=(n_4,n_5,n_6)$ respectively.

If the bulk SPT terminates at a boundary, one can dimensionally reduce the $\Theta$-term to the boundary and obtain a level-$1$ Wess-Zumino-Witten term for the $\mathrm{O}(6)$ vector $\mathbf{n}$~\cite{Bi_unpub}. The dynamics of the boundary states depends on the interactions on the boundary, and following the suggestions of Ref. [\onlinecite{Bi_unpub}] we first assume the boundary is in a gapless photon phase. The hedgehog monopoles of $\mathbf{N}$ ($-\mathbf{M}$) serve as the magnetic monopoles  (electric charges ) in the $\mathrm{U}(1)$ photon phase, and the WZW term is responsible for the statistical interaction between them. For clarity, we denote a general dyonic object by $(q,m)$ where $q$ is its electric charge and $m$ the magnetic charge.  We notice that the $(1,0)$ charges carry spin-$1/2$ under $\mathrm{SO}(3)_1$, and $(0,1)$ monopoles carry spin-$1/2$ under $\mathrm{SO}(3)_2$.

To drive the boundary into a gapped phase, we can condense $(1,0)$ or $(0,1)$. However, since they carry projective representation of the symmetry group, directly condensing them would break the symmetry. One can then condense $(2,0)$ or $(0,2)$. However, to make connection with the fSPT phase we choose to condense the double dyons $(2,2)$. Notice that a dyon $(1,1)$ is a fermion~\cite{GoldhaberPRL}, which remains deconfined after the condensation. So after the double dyons are condensed, we are left with a $\mathbb{Z}_2$ gauge theory with fermionic $\mathbb{Z}_2$ charges.

We now analyze the properties of the flux loop excitations. For this purpose, it is helpful to first go back to the photon phase, and consider adiabatically threading a $2\pi$ flux of a $\mathrm{U}(1)_1=\mathrm{SO}(2)_1$ subgroup of $\mathrm{SO}(3)_1$. Without loss of generality, we let the $\mathrm{U}(1)_1$ rotates $(n_1,n_2)$. The result of this process is a $\pi$ gauge flux loop. $(1,0)$ charges and $(1,1)$ dyons have $-1$ braiding phase with the $2\pi$ flux loop, while $m$ monopoles do not. In the photon phase, the $\pi$ flux loop will simply smear out. However, after the dyon condensation, gauge flux is quantized due to the Meissner effect. Therefore, we can identify the $2\pi$ $\mathrm{U}(1)_1$ symmetry flux loop with the $\mathbb{Z}_2$ flux loop in the gapped phase, and we postulate that its symmetry properties do not change by the condensation transition. This is highly plausible, since the double dyons (and even a single dyon) transform trivially under the symmetry group, and the condensation merely changes the energetics to impose flux quantization.

A $\mathrm{U}(1)_1$ $2\pi$ vortex loop can be represented by a vortex loop of $(n_1,n_2)$, with $\pi$ winding phase. The WZW term can be reduced to the core of the vortex loop, the result of which is a $(1+1)$-dimensional level-$1$ $\mathrm{O}(4)$ WZW term for $(n_3,n_4,n_5,n_6)$, which can be rewritten as a $\mathrm{SU}(2)_1$ WZW term~\cite{Witten84, Knizhnik84, Senthil_PRB2006}:
\begin{equation}
	S_\text{WZW}=-\frac{i}{12\pi}\int_B\di^3 y\,\varepsilon^{\mu\nu\rho}\mathrm{Tr}(U^{-1}\partial_\mu U U^{-1}\partial_\nu U U^{-1}\partial_\rho U), 
	\label{}
\end{equation}
where the $\mathrm{SU}(2)$ matrix $U=n_4+i\sum_{a=1}^3 n_a\sigma_a$.  Using the well-known Abelian bosonization formula
\begin{equation}
	U=
	\begin{pmatrix}
		e^{i\varphi} & e^{i\theta}\\
		e^{-i\theta} & e^{-i\varphi}
	\end{pmatrix},
	\label{}
\end{equation}
We can then write down the Lagrangian in terms of the bosonic fields:
\begin{equation}
	\mathcal{L}=\frac{1}{2\pi}\partial_t\varphi\partial_x\theta-\frac{v}{2\pi}[(\partial_x\varphi)^2+(\partial_x\theta)^2]+\cdots
	\label{}
\end{equation}

We can now break the symmetry down to the original $\mathbb{Z}_2$, and it is straightforward to see that
the $\mathbb{Z}_2$ symmetry action on the bosonic fields reads $\varphi\rightarrow \varphi+\pi, \theta\rightarrow \theta+\pi$, which is exactly the edge theory of a 2D $\mathbb{Z}_2$ SPT phase~\cite{LevinGu_arxiv2012, Lu_arxiv2012}. 

In summary, we have shown that the boundary of a 4D $\mathbb{Z}_2$ SPT phase can be driven into a $\mathbb{Z}_2$ topological order with fermionic $\mathbb{Z}_2$ charges, and the $\mathbb{Z}_2$ flux loops carry a nontrivial fractionalization class $[\nu]$ in $\mathcal{H}^3[\mathbb{Z}_2, \mathbb{Z}_2]$~\footnote{This result was essentially obtained in Ref. [\onlinecite{Bi_unpub}]}. It immediately implies that such a $\mathbb{Z}_2$ FTC phase can not exist in 3D systems.

Next we consider the $[\coho{w}_m]$ class.  From the perspective of dimensional reduction, such a symmetry fractionalization class corresponds to a 2D $\mathbb{Z}_2$ SET where both $e$ and $m$ anyons carry half $\mathbb{Z}_2$ symmetry charge. If such a 3D SET exists, one can consider two identical copies of them and condense the bound states of the fermionic charges from the two copies. For the 2D SET obtained from dimensional reduction, the condensation of fermion pairs seem to produce a $\mathbb{Z}_2$ SET phase where both $e$ and $m$ carry trivial $\mathbb{Z}_2$ charges. However, we notice that with both $e$ and $m$ having half $\mathbb{Z}_2$ symmetry charges, a $\mathbb{Z}_2$ symmetry flux has a topological twist equal to $\pm e^{\frac{i\pi}{4}}$~\cite{Lu_arxiv2013, BBCW, Gu_2013}. Now with two identical copies the symmetry flux penetrates both copies and the topological twist should become $(e^{i\pi/4})^2=i$. Therefore, the condensed phase actually has 2D SPT edge states (while the fractionalization class of anyons is trivial), and the 3D SET actually has a nontrivial $[\nu]$ class. In fact, what we have shown is exactly the group structure of 2D $\mathbb{Z}_2$ fSPT phases~\cite{Gu_2013}. This immediately shows that the nontrivial $[\coho{w}_m]$ is also anomalous.

\section{A construction of $\mathbb{Z}_2^T$ fermion SPT phase}
We now turn to fSPT phases with $\mathbb{Z}_2^T$ symmetry, where $\mathcal{T}^2=1$. Non-interacting fermions in 3D belong to the BDI class, which do not admit any nontrivial SPT phases~\cite{schnyder2008, kitaev2009}. However, the Gu-Wen supercohomology classification proposed a $\mathbb{Z}_2$ classification in such case, in sharp contrast to the non-interacting classification. 

\begin{figure}[t!]
	\centering
	\includegraphics[width=0.8\columnwidth]{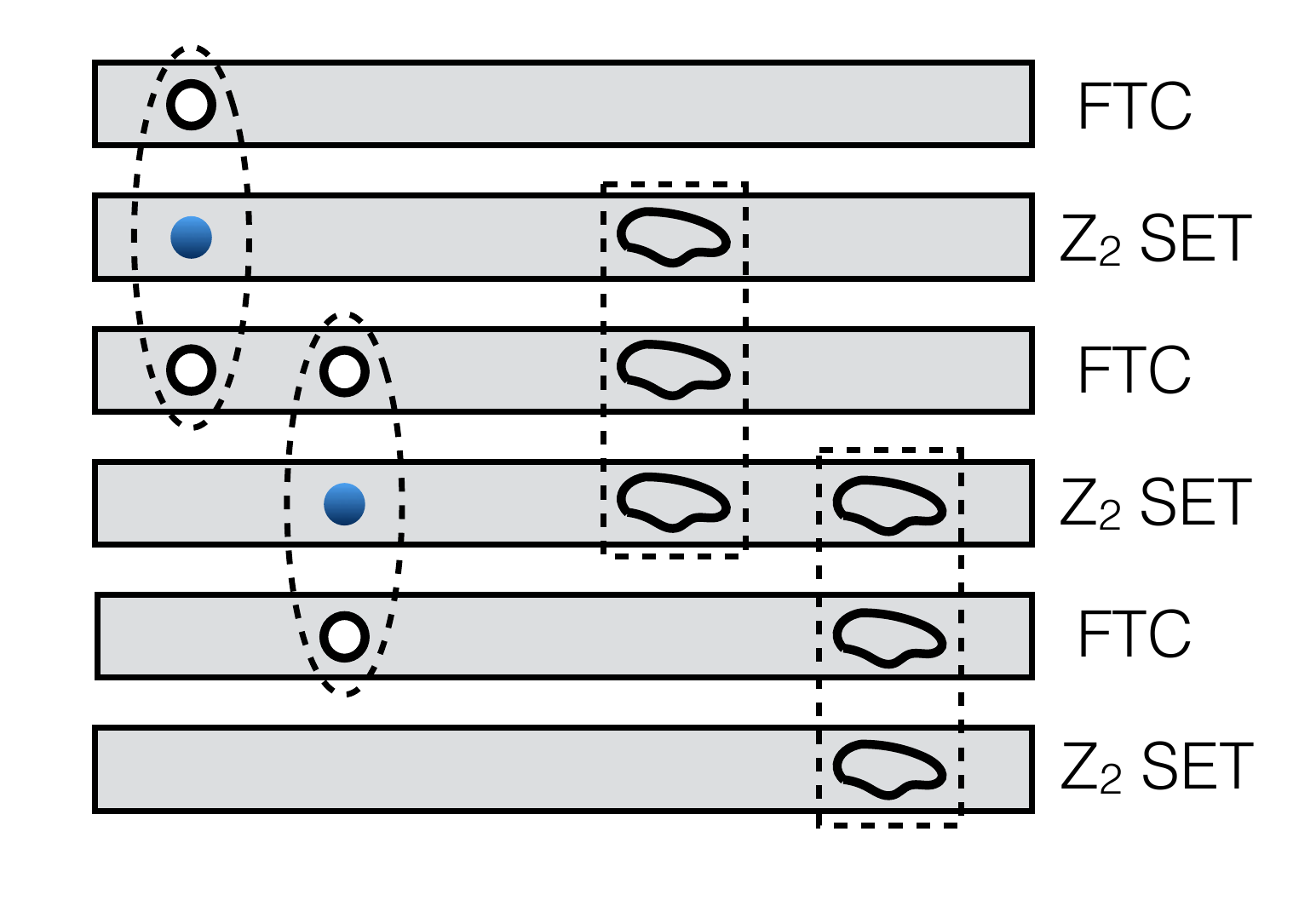}
	\caption{Illustration of the layer construction.}
	\label{fig:layer}
\end{figure}

We now propose a ``layer'' construction of $\mathbb{Z}_2^T$ fermion SPT phase by first going to 4D, inspired by similar constructions of 3D SPT phases~\cite{wang2013, JianPRX2014}. Suppose we form a stack of 3D phases as $fS_0, mS_1, fS_2, mS_3, \dots$. Here $fS$ denotes a $\mathbb{Z}_2$ FTC where $\mathbb{Z}_2$ acts completely trivially, and $mS$ denotes a $\mathbb{Z}_2$ bosonic TC characterized by a nontrivial $\mathcal{H}^3[\mathbb{Z}_2^T, \mathbb{Z}_2]$ fractionalization class on flux loops, as analyzed in Sec. \ref{sec:z2T}.  Let us proliferate $m_{2i-1}m_{2i} m_{2i+1}$, as well as $f_{2i}e_{2i+1}f_{2i+2}$ (see Fig. \ref{fig:layer} for an illustration). Here $f_{2i}$ referes to the fermionic $\mathbb{Z}_2$ charge in the $\mathbb{Z}_2$ FTC layer. One can easily check that there are no nontrivial braiding statistics between the condensed excitations, and none of them carry any nontrivial symmetry fractionalization classes, i.e. symmetry transformations on all these objects can be made completely trivial. Therefore the condensation preserves the symmetry. All the bulk excitations are confined by the condensate so we obtain a gapped phase without intrinsic topological order. However, at the boundary, $f_0$ and $m_0m_1$ remain deconfined, with the desired symmetry action:  $m_0m_1$ inherits the nontrivial $[\nu]$ fractionalization class from $mS_1$. This boundary state is exactly what we need.

We should now decouple the surface from the bulk. To do so we need to check that a) the bulk is short-ranged entangled, which we have established already  and b) The bulk does not have nontrivial symmetry-protected topological order. Since $\mathcal{H}^5[\mathbb{Z}_2^T, \mathrm{U}(1)]$ is trivial, there are no $\mathbb{Z}_2^T$ protected SPT in 4D within the cohomology classification. We should also take into account bosonic SPT states that lie beyond the cohomology classification, and there is actually such a state in 4D~\cite{KapustinSPT1, ThorngrenSPT, WenPRB2015, ZhenPRB2015}, with $\mathbb{Z}_2$-classified global gravitational anomaly on the boundary~\cite{ThorngrenSPT, KravecPRD2015}. In fact, this state does not need time-reversal symmetry for protection. The state we construct is clearly not of this kind, since if we break $\mathcal{T}$ the boundary phase is just a plain $\mathbb{Z}_2$ FTC.
Therefore, we can safely decouple the boundary 3D topological phase from the bulk. 

We need to do one extra step to realize a fermion SPT: namely we introduce physical spinless fermions into the boundary state, and condense the pair of a physical fermion and a $\mathbb{Z}_2$ charge. After the condensation, the $\mathbb{Z}_2$ fluxes are confined and in fact become $\pi$ fluxes for fermions, and we obtain a fSPT phase.  

We can see that the construction is in fact more general: for any $G$ and a $\mathcal{H}^3[G,\mathbb{Z}_2]$ fractionalization class, we can obtain a boundary $\mathbb{Z}_2$ FTC with the desired fractionalization. However, it may not be possible to decouple the boundary from the bulk if the bulk turns out to be a nontrivial SPT phase, as in the case of $G=\mathbb{Z}_2$.

\subsection{Other possibilities}
One may wonder what if we modify the construction and use a bosonic $\mathbb{Z}_2$ TC with nontrivial $[\coho{w}_m]\in \mathcal{H}^2[\mathbb{Z}_2^T, \mathbb{Z}_2]$ fractionalization, and obtain a different fermionic SPT phase. This is apparently problematic, since the resulting $\mathbb{Z}_2$ FTC, after dimension reduction becomes a 2D $eTmT$ toric code, which is known to be anomalous~\cite{senthil_3D}.  The resolution is that in this case the condensation actually breaks the symmetry: consider two loops in the condensate, $m_{2i-1}m_{2i}m_{2i+1}$ and $m_{2i+1}m_{2i+2}m_{2i+3}$. Since both of them condense, their Hopf links must also be in the condensate. However, the linking of $m_{2i+1}$ with itself is a Kramers doublet. So the condensation must break $\mathcal{T}$.

\section{Discussions}
We speculate on some future generalizations.

We have derived the classification based on localized symmetry action, however this is hardly a practical method to extract the fractionalization class given ground-state wavefunctions. The part of the classification that can be understood through dimensional reduction is relatively easier to detect, thanks to the recent progress on entanglement-based numerical methods to extract topological invariants for SET phases~\cite{ZaletelSPT, HuangPRB2014, ZaletelPSG, QiPRB2015}. It is an important problem to devise physical measurement to detect the $[\nu]$ classes that go beyond the dimensional reduction procedure.

It would be very interesting to derive a general classification of fractionalization when the $\mathbb{Z}_2$ charges have nontrivial $\mathcal{H}^2[G, \mathbb{Z}_2]$ fractionalization class~\cite{XuPRB2013}. This question is particularly relevant for the application to $\mathbb{Z}_2$ spin liquid in 3D Mott insulators, where the $\mathbb{Z}_2$ charges carry spin-$1/2$, i.e. they are spinons, as well as to develop a understanding of electronic topological insulators/superconductors using our formalism, since there gauging the fermion parity yields a $\mathbb{Z}_2$ SET where the fermionic $\mathbb{Z}_2$ charge is a Kramers doublet. For unitary $G$, it means that when $G$ is gauged, we obtain a larger gauge group $G'$ which is a nontrivial extension of $G$ by $\mathbb{Z}_2$, and one can understand the symmetry fractionalization of $\mathbb{Z}_2$ flux loops through the three-loop braiding statistics of the $G'$ gauge theory. In fact, Eq. \eqref{eqn:omega_ambiguity} already indicates a possible direction for generalization: one should refine the definition of equivalence classes for $\nu$ to mod out the additional redundancies in Eq. \eqref{eqn:omega_ambiguity}. We will leave investigation of the generalizations to future publications. 
 It has also been pointed out that SET phases with both nontrivial $[\coho{w}_e]$ and $[\nu]$ classes can be anomalous~\cite{Chen_IPAM}. It would be highly desired to develop a general theoretical framework to calculate the obstruction for 3D SET phases.

We have only considered $\mathbb{Z}_2$ topological order in this work. Generalization of our resutls to other Abelian gauge theories is fairly straightforward. What about non-Abelian gauge theories? We can take the symmetry fractionalization of gauge charges as an example. Suppose the gauge group is $H$. The gauge charges correspond to irreducible linear representations of $H$. We now need to determine the structure of projective $\mathrm{U}(1)$ phases for the gauge charges which satisfy the fusion rules. Physically, they must be given by braiding of ``Abelian'' flux loops, which correspond to the center $Z(H)$ of $H$. Therefore, the classification of symmetry fractionalizations on charges is given by $\mathcal{H}^2[G, Z(H)]$. For the gauge flux loops, we need to determine the group structure of Abelian gauge charges, or mathematically one-dimensional representations of $H$. It is known that the group of one-dimensional representations is isomorphic to the abelianization of $H$, denoted by $H^{\text{ab}}=H/[H,H]$, with $[H,H]$ being the commutator subgroup of $H$. Therefore we propose that generally the symmetry fractionalizations on a single flux loop are classified by $\mathcal{H}^3[G, H^\text{ab}]$.
In more general cases, we should also allow the possibility of symmetries permuting topological excitations. So the group cohomologies will be twisted accordingly (i.e. $G$ has a nontrivial action on the coefficients).

The symmetries studied in this work are unitary or anti-unitary on-site symmetries. An important open question is to extend the classification to space groups. In two dimensions, it has been proposed that the classification of space group fractionalization can be incorporated into a similar anyon-valued group cohomology as the on-site symmetries~\cite{essin2013, BBCW}.  It remains to see whether the $\mathcal{H}^3$ classification can be generalized in a similar way, which will be left for future works.

\section{Acknowledgement}
We thank P. Ye, C. Wang and C. Xu for enlightening conservations and Z. Bi for explaining the NLSM constructions,. We are especially grateful to X. Chen for very inspiring discussions and sharing unpublished results.

{\it Note added.} During the finalization of the manuscript we noticed a recent preprint~\cite{Fidkowski15} whose results have some overlap with ours.

\appendix

\section{Symmetry action on the edge of 2D SET phases}
\label{sec:2dsetedge}
In this section we study the symmetry action on the edge of 2D $\mathbb{Z}_2$ SET phases. We will use Chern-Simons field theory to describe the bulk topological phase, and the edge theory is a Luttinger liquid. For our purpose, it is sufficient to take the K matrix to be $K=2\sigma_x$, and the edge theory given by
\begin{equation}
	\mathcal{L}=\frac{1}{\pi}\partial_t\phi\partial_x\theta - v_1(\partial_x\theta)^2-v_2(\partial_x\phi)^2+\cdots
	\label{}
\end{equation}
Here $\phi$ and $\theta$ are both $2\pi$-periodic bosonic fields. Without loss of generality, we can assume $e^{i\phi}$ ($e^{i\theta}$) corresponds to a $e$($m$) anyon. We consider a SET phase with $\mathbb{Z}_2$ symmetry, and assume the corresponding symmetry transformation on the edge is given by
\begin{equation}
	U_\mb{g}\phi U_\mb{g}^{-1}=\phi, U_\mb{g}\theta U_\mb{g}^{-1}=\theta+\frac{n\pi}{2}.
	\label{}
\end{equation}
Here $\mb{g}$ is the nontrivial element of $\mathbb{Z}_2$ symmetry group, and $n\in\mathbb{Z}$.

From the symmetry transformation we can easily find the following form for $U_\mb{g}$:
\begin{equation}
	U_\mb{g}=\exp\left( \frac{in}{2}\int\di x\, \partial_x\phi\right).
	\label{}
\end{equation}
Therefore, from $U_\mb{g}^2=\Omega(\mb{g,g})$ we find
\begin{equation}
	\Omega(\mb{g,g})=\exp\left( {in}\int\di x\, \partial_x\phi\right).
	\label{}
\end{equation}
Physically we can interpret $\Omega(\mb{g,g})$ as $n$ copies of closed string operators of $e$.

\section{Derivation of the $3$-cocycle condition}
\label{sec:3cocycle}
In this section we derive the $3$-cocycle condition for $\nu(\mb{g,h,k})$. The derivation follows Ref. [\onlinecite{ElseSPT}], but we also include anti-unitary symmetries. We introduce a $\mathbb{Z}_2$ grading $q(\mb{g})=0,1$, such that $q(\mb{g})=0/1$ means $\mb{g}$ represents a unitary/anti-unitary operation.

First we localize the global symmetry operations to the boundaries. Notice that because complex conjugation is a global operation on the wavefunction, it is unclear how to localize $K$. We then define
\begin{equation}
	R_\mb{g}\ket{\psi}=\prod_{\partial M}U_\mb{g} K^{q(\mb{g})}\ket{\psi}.
	\label{}
\end{equation}
Notice that $U_\mb{g}$'s are unitary operators. Demanding that $R_\mb{g}$ forms a linear representation,
\begin{equation}
	U_\mb{g} K^{q(\mb{g})} U_\mb{h} K^{q(\mb{g})}=U_\mb{gh}.
	\label{}
\end{equation}
We can then restrict $U_\mb{g}$ to an interval $I$ of the boundary, and as before
\begin{equation}
	U^I_\mb{g} K^{q(\mb{g})} U^I_\mb{h} K^{q(\mb{g})}=\Omega(\mb{g,h})U^I_\mb{gh}.
	\label{}
\end{equation}

To derive associativity relation of $\Omega$, we consider $U_\mb{g}^I K^{q(\mb{g})}[U^I_\mb{h}K^{q(\mb{h})} U^I_\mb{k} K^{q(\mb{h})}]K^{q(\mb{g})}$ which can be evaluated in two ways:
\begin{widetext}
	\begin{equation}
		\begin{split}
		U_\mb{g}^I K^{q(\mb{g})}[\Omega(\mb{h,k})U^I_\mb{hk}]K^{q(\mb{g})}&=^{U_\mb{g}^I}\!\big[K^{q(\mb{g})}\Omega(\mb{h,k})K^{q(\mb{g})}\big]\,U_\mb{g}^I K^{q(\mb{g})}U^I_\mb{hk}K^{q(\mb{g})}=^{U_\mb{g}^I}\!\big[K^{q(\mb{g})}\Omega(\mb{h,k})K^{q(\mb{g})}\big]\Omega(\mb{g,hk})W_{\mb{ghk}}^I\\
	&=\Omega(\mb{g,h})U_\mb{gh}^I K^{q(\mb{gh})} U^I_\mb{k} K^{q(\mb{h})}K^{q(\mb{gh})}=\Omega(\mb{g,h})\Omega(\mb{gh,k})U_\mb{ghk}^I.
		\end{split}
	\label{}
\end{equation}
Therefore, we have
\begin{equation}
	\Omega(\mb{g,h})\Omega(\mb{gh,k})=^{U_\mb{g}^I}\!\big[K^{q(\mb{g})}\Omega(\mb{h,k})K^{q(\mb{g})}\big]\Omega(\mb{g,hk}).
	\label{}
\end{equation}

The restriction only satisfies the relation up to a phase:
\begin{equation}
	\Omega_a(\mb{g,h})\Omega_a(\mb{gh,k})=\nu(\mb{g,h,k})^{U_\mb{g}^I}\!\big[K^{q(\mb{g})}\Omega_a(\mb{h,k})K^{q(\mb{g})}\big]\Omega_a(\mb{g,hk}).
	\label{}
\end{equation}

To derive the $4$-cocycle condition for $\nu(\mb{g,h,k})$, consider $\Omega_a(\mb{g,h})\Omega_a(\mb{gh,k})\Omega_a(\mb{ghk,l})$:
\begin{equation}
	\begin{split}
	\Omega_a&(\mb{g,h})\Omega_a(\mb{gh,k})\Omega_a(\mb{ghk,l})\\
	&=\nu(\mb{gh,k,l})\Omega_a(\mb{g,h})^{U_\mb{gh}^I}\!\big[K^{q(\mb{gh})}\Omega_a(\mb{k,l})K^{q(\mb{gh})}\big]\Omega_a(\mb{gh,kl})\\
	&=\nu(\mb{gh,k,l})^{\Omega_a(\mb{g,h})U_\mb{gh}^I}\!\big[K^{q(\mb{gh})}\Omega_a(\mb{k,l})K^{q(\mb{gh})}\big]\Omega_a(\mb{g,h})\Omega_a(\mb{gh,kl})\\
	&=\nu(\mb{gh,k,l})\nu(\mb{g,h,kl})^{\Omega_a(\mb{g,h})U_\mb{gh}^I}\!\big[K^{q(\mb{gh})}\Omega_a(\mb{k,l})K^{q(\mb{gh})}\big] ^{U_\mb{g}^I}\!\big[K^{q(\mb{g})}\Omega_a(\mb{h,kl})K^{q(\mb{g})}\big]\Omega_a(\mb{g, hkl}).
	\end{split}
	\label{}
\end{equation}

Proceeding in a different order,
\begin{equation}
	\begin{split}
	\Omega_a&(\mb{g,h})\Omega_a(\mb{gh,k})\Omega_a(\mb{ghk,l})\\
	&=\nu(\mb{g,h,k})^{U_\mb{g}^I}\!\big[K^{q(\mb{g})}\Omega_a(\mb{h,k})K^{q(\mb{g})}\big]\Omega_a(\mb{g,hk})\Omega_a(\mb{ghk,l})\\
	&=\nu(\mb{g,h,k})\nu(\mb{g,hk,l})^{U_\mb{g}^I}\!\big[K^{q(\mb{g})}\Omega_a(\mb{h,k})K^{q(\mb{g})}\big]^{U_\mb{g}^I}\!\big[K^{q(\mb{g})}\Omega_a(\mb{hk,l})K^{q(\mb{g})}\big]\Omega_a(\mb{g,hkl})\\
	&=\nu(\mb{g,h,k})\nu(\mb{g,hk,l})^{U_\mb{g}^I}\!\big[K^{q(\mb{g})}\Omega_a(\mb{h,k})\Omega_a(\mb{hk,l})K^{q(\mb{g})}\big]\Omega_a(\mb{g,hkl})\\
	&=\nu(\mb{g,h,k})\nu(\mb{g,hk,l})^{U_\mb{g}^I}\!\big\{K^{q(\mb{g})}\nu(\mb{h,k,l})^{U_\mb{h}^I}\!\big[K^{q(\mb{h})}\Omega_a(\mb{k,l})K^{q(\mb{h})}\big]\Omega_a(\mb{h,kl})K^{q(\mb{g})}\big\}\Omega_a(\mb{g,hkl})\\
	&=\nu(\mb{g,h,k})\nu(\mb{g,hk,l})K^{q(\mb{g})}\nu(\mb{h,k,l})K^{q(\mb{g})}\, ^{U_\mb{g}^I}\!\big\{K^{q(\mb{g})}\, ^{U_\mb{h}^I}\!\big[K^{q(\mb{h})}\Omega_a(\mb{k,l})K^{q(\mb{h})}\big]\Omega_a(\mb{h,kl})K^{q(\mb{g})}\big\}\Omega_a(\mb{g,hkl})\\
	&=\mb{g}[\nu(\mb{h,k,l})]\nu(\mb{g,h,k})\nu(\mb{g,hk,l}) ^{U_\mb{g}^IK^{q(\mb{g})}U_\mb{h}^IK^{q(\mb{g})}}\!\big[K^{q(\mb{gh})}\Omega_a(\mb{k,l})K^{q(\mb{gh})}\big]^{U_\mb{g}^I}\!\big[K^{q(\mb{g})}\Omega_a(\mb{h,kl})K^{q(\mb{g})}\big]\Omega_a(\mb{g,hkl})\\
	&=\mb{g}[\nu(\mb{h,k,l})]\nu(\mb{g,h,k})\nu(\mb{g,hk,l}) ^{\Omega(\mb{g,h})U_\mb{gh}^I}\!\big[K^{q(\mb{gh})}\Omega_a(\mb{k,l})K^{q(\mb{gh})}\big]^{U_\mb{g}^I}\!\big[K^{q(\mb{g})}\Omega_a(\mb{h,kl})K^{q(\mb{g})}\big]\Omega_a(\mb{g,hkl}).
\end{split}
\end{equation}
Here $\mb{g}[x]=K^{q(\mb{g})}xK^{q(\mb{g})}$.

We need to argue that
\begin{equation}
	^{\Omega(\mb{g,h})U_\mb{gh}^I}\!\big[K^{q(\mb{gh})}\Omega_a(\mb{k,l})K^{q(\mb{gh})}\big]=^{\Omega_a(\mb{g,h})U_\mb{gh}^I}\!\big[K^{q(\mb{gh})}\Omega_a(\mb{k,l})K^{q(\mb{gh})}\big].
	\label{}
\end{equation}
\end{widetext}
This follows from the assumption that $\Omega_a$ are localized near the end point $a$, as well as that $K$ maps local operators to local operators.

So we have derived the $4$-cocycle condition:
\begin{equation}
	\mb{g}[\nu(\mb{h,k,l})]\nu(\mb{g,h,k})\nu(\mb{g,hk,l})=\nu(\mb{gh,k,l})\nu(\mb{g,h,kl}).
	\label{}
\end{equation}

\bibliography{spt}

\end{document}